\newcommand{\NATP}[3]{Nat.\ Phys.\ {\bf A#1},\ #2 (#3)}

\newcommand{\PLA}[3]{Phys.\ Lett.\ A\ {\bf #1},\ #2 (#3)}

\newcommand{\PR}[3]{Phys.\ Rev.\ {\bf #1},\ #2 (#3)}
\newcommand{\PRL}[3]{Phys.\ Rev.\ Lett.\ {\bf #1},\ #2 (#3)}

\newcommand{\OC}[3]{Opt.\ Commun.\ {\bf #1},\ #2 (#3)}

\newcommand{\PRA}[3]{Phys.\ Rev.\ A\ {\bf #1},\ #2 (#3)}

\newcommand{\JPA}[3]{J.\ Phys.\ A:\ Math.\ Gen.\ {\bf #1},\ #2 (#3)}

\newcommand{\EPJD}[3]{Eur.\ Phys.\ J.\ D\ {\bf #1},\ #2 (#3)}

\newcommand{\AIP}[3]{AIP.\ Conf.\ Proc. \  {\bf #1},\ #2 (#3)}
\documentclass[aps,showkeys,showpacs,amsmath,amssymb,twocolumn,superscriptaddress,prl]{revtex4}
\usepackage{amsmath}
\usepackage{amssymb}
\usepackage{dcolumn}
\usepackage{graphicx}
\usepackage{longtable}
\usepackage{subfigure}
\usepackage{color}
\newtheorem{thm}{Theorem}
\newtheorem{lem}{Lemma}
\begin{document}
\title{In how many ways can quantum information be split ?}

\author{S. Muralidharan}
\affiliation{School of EPS, Heriot-Watt University, Edinburgh, EH144AS, United Kingdom}
\author{S. Karumanchi}
\affiliation{Birla Institute of Technology and Science, Pilani, Rajasthan- 333031, India}
\author{S. Narayanaswamy}
\affiliation{Poornaprajna Institute of Scientific Research, Sadashivnagar, Bangalore 560 080, India}
\author{R. Srikanth}
\affiliation{Poornaprajna Institute of Scientific Research, Sadashivnagar, Bangalore 560 080, India}
\affiliation{Raman Research Institute, Sadashiva Nagar, Bangalore 560012, India}

\author{P. K. Panigrahi}
\email{pprasanta@iiserkol.ac.in}
\affiliation{Indian Institute of Science Education and Research - Kolkata, Mohanpur - 741252, India}

\begin{abstract}
We establish a theoretical understanding of the entanglement properties of a physical system that mediates a quantum information splitting protocol. We quantify the different ways in which an arbitrary $n$ qubit state can be split among a set of $k$ participants using a $N$ qubit entangled channel, such that the original information can be completely reconstructed only if all the participants cooperate.  Based on this quantification, we show how to design a quantum protocol with minimal resources and define the splitting efficiency of a quantum channel which provides a way of  characterizing entangled states based on their usefulness for such quantum networking protocols. 

\end{abstract}
\pacs{03.67.Hk, 03.65.Ud}
\keywords{Quantum information splitting, Teleportation}
\maketitle

\textit{Introduction-}Splitting and sharing of secret information among a group of parties such that none of them can completely reconstruct the secret information by themselves is a common requirement in financial and defence sectors \cite{Bruce}. Recently, it has been found that  certain aspects of quantum mechanics such as entanglement  \cite {EPR} can be effectively used for  the splitting and sharing of secret information which can be either `classical' (bits) or `quantum' (qubits) \cite{Gotit}. 
Sharing of quantum information among a group of parties such that none of them can reconstruct the unknown
information completely by operating on their own share is usually referred to as ”Quantum Information Splitting” (QIS).  Starting from the seminal work of Hillery \it  {et al.} \normalfont\cite{Hillery}  who used a $N$-qubit Greenberger-Horne-Zeilinger (GHZ) state for the splitting up of an unknown single qubit among many parties, several entangled states were found to be useful for this purpose \cite{Zheng, Sreraman1, Sreraman2, Sreraman3, Sreraman4, Sreraman5}. Among these, `graph-states' are of particular interest as they have been experimentally realized upto six-qubits \cite{Damian, cluster}. A schematic view of a standard QIS protocol is presented in Fig. 1. 
\begin{figure}[h]
\begin{center} 
\includegraphics[height=2.5in,width=1.5in,angle=270]{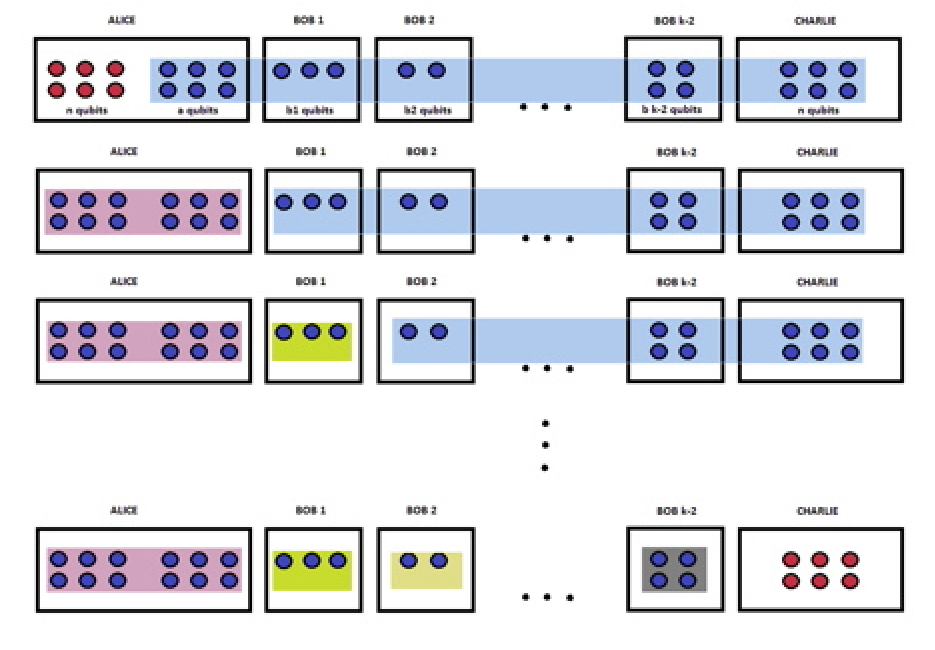} 
\caption{\small \sl Alice, Bobs and Charlie share an entangled channel. Alice performs a joint-measurement on the secret and her share of the entangled channel. Each party performs a measurement on his/her qubits and subsequently, the secret is split among other parties.} 
\end{center} 
\end{figure}  
\\In a realistic situation, for the splitting of an arbitrary number of qubits, many parties need to be in an entangled quantum network. 
Keeping in mind the complexity of the multipartite entangled system, 
there will arise more than one way of splitting and sharing of secret information, 
given a fixed number of parties. For instance, if a dealer want to split an unknown single qubit state $|\psi_1\rangle = \alpha|0\rangle+\beta|1\rangle$ ($\alpha, \beta \in C, |\alpha|^2 + |\beta|^2=1 $) among two parties, using a symmetric four-qubit entangled GHZ state, $|\phi_{2345}\rangle$  as a resource \cite{Hillery}, this could be done in two different ways as their combined state could be written as
\begin{eqnarray}
|\psi_1\rangle \otimes |\phi_{2345}\rangle = \sum_{i=1}^{4} \sum_{j=1}^{2} (|\psi_{12}\rangle_{i} \otimes |\phi_{34}\rangle_{j} \otimes |\phi_{5}\rangle_{j}) \\ \nonumber
= \sum_{i=1}^{4} \sum_{j=1}^{2} (|\psi_{123}\rangle_{i} \otimes |\phi_{4}\rangle_{j} \otimes |\phi_{5}\rangle_{j}). 
\end{eqnarray}
In the first case, $|\psi_{12}\rangle_{i}$'s forms four orthogonal two-qubit measurement outcomes of the dealer and $|\phi_{34}\rangle_{j}$'s forms  two two-qubit
orthogonal measurement outcomes of the intermediate party. In the second case, $|\psi_{123}\rangle_{i}$'s forms the  four 
orthogonal three qubit measurement outcomes of the dealer and $|\phi_{4}\rangle_{j}$'s forms the two, single qubit measurement outcomes
of the intermediate party.  The dealer can split the outcome of his/her measurement using a (2,2)-threshold classical secret sharing (CSS). 
If the participants co-operate, the receiver can covert $|\phi_{5}\rangle_{j}$ to $|\psi_1\rangle$ by applying an appropriate unitary operation. 

Note that all distributions of qubits among the parties don't yield successful QIS protocols. Further, one needs to choose an entangled channel with appropriate number of qubits  given the number of qubits in the secret and the number of parties among whom this secret information has to be split. In fact, 
 it has been observed that a four-qubit linear cluster state could not be used for the QIS of an unknown two-qubit state while a five qubit linear cluster state could be used for this purpose \cite{Sreraman1}.
In this Letter, we study the entanglement properties of the physical system which mediates this QIS protocol and quantify the
number of ways in which quantum information could be split among a given number of parties. This quantification provides a recipe for the design a QIS protocol with minimal entangled resources. Based on this study we define \textit{splitting efficiency} of a quantum channel, which provides an alternative method for characterizing quantum states based on their usefulness for QIS. The introduced model seems to be naturally applicable for many quantum networking protocols \cite{multiparty} that relies on entanglement and teleportation \cite{Ben}.  We establish the quantification first and then the efficiency of QIS in the later part of the letter.\\
\textit{Protocol count-}Following theorems yield the protocol count for the QIS of an unknown $n$-qubit state $|\psi_{n}\rangle = \displaystyle\sum\limits_{i_{1},...i_{n}=0}^{1} \alpha_{i_{1},...i_{n}}|i_{1},...i_{n}\rangle$, where $\alpha_{i_{1},...i_{n}}\in C$ and $\sum |\alpha_{i_{1},...i_{n}}|^2 = 1$.
\begin{thm}
\label{thm:capacity}
If Alice, Bob(s) and Charlie share an $N$ qubit entangled state and Alice
has a  arbitrary $n$ qubit  state $|\psi_n\rangle$ that she  wants the Bobs
and Charlie  to share, then  Alice needs to  possess a minimum  of $n$
qubits for this purpose.
\end{thm}
\textit{Proof: }We conflate all Bobs and Charlie into a single agent Dolly, equipped with Hilbert space ${\cal H}_D$, which is the tensor product of the Hilbert spaces of the Bobs and Charlie. Information splitting can be considered as a quantum teleportation from Alice to Dolly. Teleportation with unit fidelity will not be deterministically possible unless maximal entanglement exists between Alice and Dolly \cite{agpa}.
 By virtue of Schmidt decomposition for a bipartite pure state, Dolly's density operator will be maximally mixed in a $2^n$-dimensional {\it subspace} of ${\cal H}_D$. If we  let Alice possess  $m$ qubits in  the entangled  quantum network then  it can be proved that $m \geq  n$ from a quantum  encryption perspective as follows:  After Alice's joint measurement in
 ${\cal   H}_x  \otimes   {\cal   H}_A$,  but   before  her   classical  communication  to Dolly,  the  no-signalling theorem  demands that  Dolly's
 density  operator  should  not  have  changed, i.e.,  it  must  remain  maximally mixed within the relevant subspace. Here, ${\cal  H}_x$ and ${\cal  H}_A$ refer  to the  respective  Hilbert  spaces  of  the unknown  secret  information  and  Alice's part of the entangled network respectively. On the other hand,
we know by the way deterministic teleportation works (for the considered, maximally
entangled states) that Dolly's state has become: 
 
 \begin{equation}
 {\bf    T}:    |\psi\rangle    \longrightarrow   \sum_{j=1}^{P}    U_j
 |\psi\rangle\langle \psi| U^\dag_j 
 \label{eq:encrypt} 
 \end{equation}

 Alice's  classical communication  will be  the number  $j$  which will
 allow Dolly to apply operation $U_j$ that restores her object's state to
 $|\psi\rangle$.   We   require  the   minimal   number   $P$  in   Eq.
 (\ref{eq:encrypt})   such   that   for   an  arbitrary   input   state
 $|\psi\rangle$, we  obtain ${\bf T}(|\psi\rangle) =  {\bf I}/D$, where
 ${\bf   I}$   is  the   unit   matrix   and   $D=2^n$  is   dim(${\cal
 H}_B$). According to Ref.  \cite{mosca}, which provides a protocol for
 classically encrypting a quantum state,  $P = D^2$. This in turn means
 that Alice's classical communication must be at least $\log(D^2) = 2n$ bits long. In
 turn this  means that the outcome of Alice's measurement,
 which could be encoded into $(m+n)$ bits, must satisfy $m+n \ge 2n$, or, $m \ge n$, as required. \hfill
$\blacksquare$
It is worth mentioning that Theorem 1 also follows as a consequence of the Choi-Jamiolkowski Isomorphism \cite{CJ}. 
\begin{thm} 
\label{lem:capacity2}
It is necessary for the recipient's system to be in a maximally mixed state, 
but not for that of any intermediate party {\bf P}.
\end{thm} 
\textit{Proof: }By application of the Agrawal-Pati theorem \cite{agpa}, the $n$ qubits
with the recipient should be maximally mixed. The reduced density operator
for any intermediate party {\bf P}, however, need not be maximally mixed
in its support. To see this, consider the scenario when three parties Alice, Bob and Charlie share an entangled state of the form
$|\zeta\rangle \equiv \cos\theta|{+}\rangle_B|\psi^-\rangle_{AC} + 
\sin\theta|{-}\rangle_B|\psi^+\rangle_{AC}$, where, $|\pm\rangle  = \frac{1}{\sqrt{2}} (|0\rangle \pm |1\rangle)$ and $\theta \in [0, \pi/2]$.
This is manifestly non-maximally entangled for $\theta \neq \pi/4 $. 
Since Alice's and Bob's measurements commute,
and Bob does not use Alice's classical communication, Bob might as well
measure first. If he does to obtain the outcome $|\pm\rangle$ clearly Alice is able to deterministically teleport to Charlie
with fidelity 1. This is alternatively confirmed by seeing that Alice's and Charlie's reduced density operator
for $|\zeta\rangle$ is maximally mixed (in its support). These observations
give us Theorem \ref{lem:capacity2}.   
 \hfill $\blacksquare$
\begin{lem} The maximum number of protocols one can  construct 
for this purpose is $(N-2n)$.
\label{lem:capacity}
\end{lem}

\textit{Proof: }We  let the third person (say Charlie)  have the last $n$
qubits, on which he will apply a suitable local unitary transformation
and reconstruct  the unknown $n$ qubit  information. Therefore Charlie
will  possess, $(N -  n +  1)^{\rm  th}$ qubit  to  the $N^{\rm  th}$
qubit. Now,  the first $(N -  n)$ qubits need to  be distributed among
Alice   and  Bob.   This  would   correspond  to 
  $(N  -   n   -  1)$
protocols.  However, from Theorem 1, all  the  protocols, in
which  Alice possesses  less than  $n$ qubits  fail. Hence,  the total
number of protocols that one can construct is $(N - 2n)$. For at least
one  protocol   to  work   out,  we  have   $N  \geq   (2n+1)$, yielding a recipe for the design of a QIS protocol with 
minimal entangled resources. \hfill
$\blacksquare$

{\bf Corollary.}  By substituting $N = 4$ and $n = 2$ in this formula,
we can  deduce that four  qubit states cannot  be used for the  QIS of an unknown two qubit state
$|\psi_2\rangle$ . This  shall be illustrated below as  follows. We let
Alice possess  the unknown two  qubit state $|\psi_2\rangle$
and qubit  1, Bob possess  qubit 2 and  Charlie 3,4 in the  four qubit
cluster state \cite{Sreraman1}, $|C_4\rangle = \frac{1}{2}(|0000\rangle + |0110\rangle + |1001\rangle - |1111\rangle)_{1234} $
respectively. Alice can perform a three partite measurement on $|\psi_2\rangle$ as in the above protocol.
 For instance, if Alice performs a  three-qubit measurement and obtains the outcome $\frac{1}{2} ( |000\rangle+|100\rangle+|011\rangle-|111\rangle)$, then, the Bob-Charlie system collapses to $\alpha_{00}(|000\rangle+|110\rangle)+\alpha_{01}(|000\rangle+|110\rangle) +\alpha_{10}(|001\rangle-|111\rangle)+\alpha_{11}(|001\rangle-|111\rangle)$. 
However, from  the above state one cannot  obtain $|\psi_2\rangle$, by performing another  measurement on some of its qubits to  get $|\psi\rangle_2$.   This could be seen more clearly as follows, 
If Bob performs a single particle measurement and obtains the 
outcomes $\frac{1}{\sqrt{2}} (|0\rangle \pm |1\rangle)$, 
then Charlie's state collapses to  $\alpha_{00}(|00\rangle+|10\rangle)+\alpha_{01}(|00\rangle+|10\rangle)+ \alpha_{10}(|01\rangle-|11\rangle)+\alpha_{11}(|01\rangle-|11\rangle)$. If Charlie can apply an operation $U$ and obtain $|\psi_2\rangle$, it can be clearly seen using simple matrix algebra that there exists no unitary operator $U$,
 which is independent of the coefficients that carries out the task. Since, the coefficients of $|\psi_2\rangle$ are unknown, we conclude that this protocol fails illustrating the usefulness of the theorem. \\ 
We define ${\bf P}_k(N)$ as the number theoretic partitions, ${\bf Q}_k(N)$ as the ordered partition of $N$ restricted to $k$ terms, $C$ to be the binomial coefficient and in general ${\bf Q}_k(N) \ge {\bf P}_k(N)$. On account of Theorem \ref{thm:capacity}, the dealer must have at least $n$ qubits to initiate the QIS protocol and the receiver must have precisely $n$ qubits (ignoring extraneous qubits) to be able to reconstruct the secret. 
Depending on the pre-shared entanglement chosen, the protocol may be
symmetric between the Bobs, in which case interchanging two 
Bobs is equivalent or completely asymmetric between all of them.

\begin{thm}
If $k$ $(3 \leq k \leq N-2n+2)$ parties share  an $N$ qubit entangled state and the
first party  has an  arbitrary $n$ qubit  state that he/she  wants the
remaining  members to  share,  then the  maximum  number of  protocols
that can be constructed for this purpose is 
$\sum_{j=k-2}^{N-2n} {\bf P}_{k-2}(j)$ in the symmetric case, and
bounded above by $\sum_{j=k-2}^{N-2n} {\bf Q}_{k-2}(j) =
{^{N-2n}}C_{k-2}$ in the general case.
\label{thm:splits}
\end{thm}

{\bf Proof.} The minimum number of qubits with the Bobs is $(k-2)$ and the
maximum number, $(N-2n)$. Fix the number of qubits
 with all Bobs together to
be $j$. The number of protocols in the symmetric case is the number of ways
$j$ can be partitioned into $k-2$ slots (with each slot having at least
1 entry), which is ${\bf P}_{k-2}(j)$. Summing over all possible $j$'s gives the 
total number of protocols in the symmetric case. 
If the state is such that
each Bob is inequivalent to any other, then clearly 
the number of protocols is $\sum_{j=k-2}^{N-2n} {\bf Q}_{k-2}(j) = \sum_{j=k-2}^{N-2n} {^{j-1}}C_{k-3}
= {^{N-2n}}C_{k-2}$, for it can be shown that ${\bf Q}_l(m) = {^{m-1}}C_{l-1}$. In general,
this is an upper bound on the number of protocols, as there will be
partial symmetry among the Bobs. In order for atleast one protocol to work out we have 
 $N-2n \geq k-2$. \hfill $\blacksquare$\\

While Theorem \ref{thm:capacity} specifies the threshold size of Alice's system to be useful for QIS of completely unknown quantum states,  one may ask about
the threshold size required for the QIS of specific types of quantum states of the form $|\phi_{n}\rangle = \alpha {|0\rangle}^{\otimes n} + \beta {|1 \rangle}^{\otimes n}$. 
This leads us to Theorem 4. 
\begin{thm} If Alice, Bob and Charlie share an $N$-qubit 
entangled state and Alice has an (entangled) $n$-qubit 
entangled state of the form 
$|\phi_{n}\rangle = \alpha {|0\rangle}^{\otimes n} + \beta {|1 \rangle}^{\otimes n}$ that she wants
Bob and Charlie to share, then Alice needs to possess only 
one qubit for this purpose.
\label{thm:qis}
\end{thm}
{\bf Proof:}  
Consider a scenario where Alice, Bob and Charlie share a symmetric $N$-qubit GHZ state of the form, 
$ |\phi_{GHZ}\rangle= \frac{1}{\sqrt{2}}(|0\rangle^N + |1\rangle^N)$, where $N = 2+n$, with the first qubit with Alice,
second with Bob, and the remaining with Charlie. Alice can perform a $(n + 1)$ particle measurement on her qubits and conveys 
its outcome to Charlie (or Bob) using $(n + 1)$ classical bits. Her outcome and the corresponding Bob-Charlie states are:
\begin{center}
\begin{tabular}{|c|c|} \hline
	Alice's Outcome 	& Bob-Charlie State \\ \hline
	$|\psi_1\rangle$	& $\alpha {|0\rangle}^{\otimes (n + 1)} + \beta {|1\rangle}^{\otimes (n + 1)}$ \\
	$|\psi_2\rangle$	& $\alpha {|0\rangle}^{\otimes (n + 1)} - \beta {|1\rangle}^{\otimes (n + 1)}$ \\
	$|\psi_3\rangle$	& $\alpha {|1\rangle}^{\otimes (n + 1)} + \beta {|0\rangle}^{\otimes (n + 1)}$ \\
	$|\psi_4\rangle$	& $\alpha {|1\rangle}^{\otimes (n + 1)} - \beta {|0\rangle}^{\otimes (n + 1)}$ \\ \hline
\end{tabular} 
\end{center}
\noindent
where $\sqrt{2}|\psi_{1, 2}\rangle = {|0\rangle}^{\otimes (n + 1)} \pm  {|1\rangle}^{\otimes (n + 1)}$ and
$\sqrt{2}|\psi_{3, 4}\rangle = {|0\rangle}^{\otimes n} \otimes |1\rangle \pm {|1\rangle}^{\otimes n} \otimes |0\rangle$
form mutual orthogonal outcomes of measurement. Using Alice's 2-bit communication, Charlie (or Bob) can 
apply the single-qubit Pauli 
operation $I, Z, X$ or $Y$ to bring their entangled state to
the form $\alpha|0\rangle^{\otimes (n+1)} + \beta|1\rangle^{\otimes (n+1)}$.
By measuring each of his/her qubit in a suitable basis to obtain the outcome  $|\pm\rangle$ and communicating
his 1-bit outcome to Charlie, the latter can reconstruct $|\phi_n\rangle$. If Alice's 2-bit outcome is known, Bob or Charlie possess
partial information about $\alpha$ or $\beta$, and hence about
$|\phi_n\rangle$. As before, this partial information can be eliminated by
having Alice split her 2-bit information between Bob
and Charlie using a (2,2)-threshold CSS scheme. 
\hfill $\blacksquare$

\begin{lem} The number of protocols one can construct for this purpose is $^{N - 2n}{\rm C}_{k - 2} + n - 1$. 
\label{lem:restricted}
\end{lem}
{\bf Proof.}
It was shown in Lemma 1 that one can construct $(N - 2n)$ protocols for $k=3$ in the case where Alice had to possess at least $n$ qubits. But, for the present case the protocols in which Alice possesses less than $n$ qubits also work for $n \geq 2$. Given that the recipient has $n$ qubits, the remaining $(N - n)$ qubits is to be split among the remaining parties, this can be done in ($N - n - 1$) ways. Following a similar line of reasoning as given in Theorem 3, we can see that for $k\geq3$, one can construct  $^{N - 2n}{\rm C}_{k - 2} + n - 1$ protocols for this purpose. 
\hfill $\blacksquare$\\
\textit{Efficiency of Information Splitting- }The above theorems naturally lead to questions about the efficiency of a quantum channel for QIS.
We define the "splitting efficiency" ($\eta$) of a quantum channel as,
\begin{equation}
\eta = \frac{n_{0}}{n_{\max}}\frac{ \sum _{n = 1}^{n_{\max}}n \zeta_{n}}{{ \sum _{n = 1}^{n_{\max}}n \zeta '_{n}}},
\end{equation}
where $\zeta_{n}$ refers to the number of protocols that can be constructed for splitting up of $|\psi_n\rangle$ among $k$ parties for a given entangled channel, $\zeta'_{n}$ refers to the maximum number of protocols that can be constructed for splitting up of  $|\psi_n\rangle$ among $k$ parties as given by Theorem 3.
Here, $n_{\max} = \lfloor \frac{N - k + 2}{2} \rfloor$ does not depend on the particular channel, but on $N$ and $k$ only and $n_{0}$ is 
the largest size of a secret (in qubit units) that can be split with $N$ qubit entangled channel among $k$ parties. To illustrate this, we consider a maximally entangled $N$ qubit GHZ state given by, $|\phi_{GHZ}\rangle$ From Theorems 1 and 2, we can note that   $|\phi_{GHZ}\rangle$ can be used for the QIS of only unknown single qubit state $|\psi_1\rangle$. From Eq. (3),  we can calculate its the splitting efficiency as  $\eta_{GHZ} = (1/n_{\max})$. For a $N$ qubit linear cluster state \cite{linear}, which can be used for the QIS of both arbitrary single and two qubit systems, the splitting capacity is given by, $\eta_{lc} = ({1}/{n_{\max}}) \ {\rm if} \ N =3, 4 $ and $(2/{n_{\max}})  \ {\rm if} \ N \geq 5$. As it can be seen, $\eta_{lc} > \eta_{GHZ}$ for $N>4$ showing that linear cluster states are more useful than GHZ states for QIS. However, $2$X$N$ cluster states ( Box states) \cite{Box} exhibit even more interesting properties than the linear cluster
states as all the possible protocols work out for QIS among three parties. The splitting capacity of box states is given by, $\eta_{Box} = 1$. A detailed comparison of the splitting efficiencies for these entangled channels is shown in Fig. 2. 

\begin{figure}
\includegraphics[height=1.5in,width=2in,angle=0]{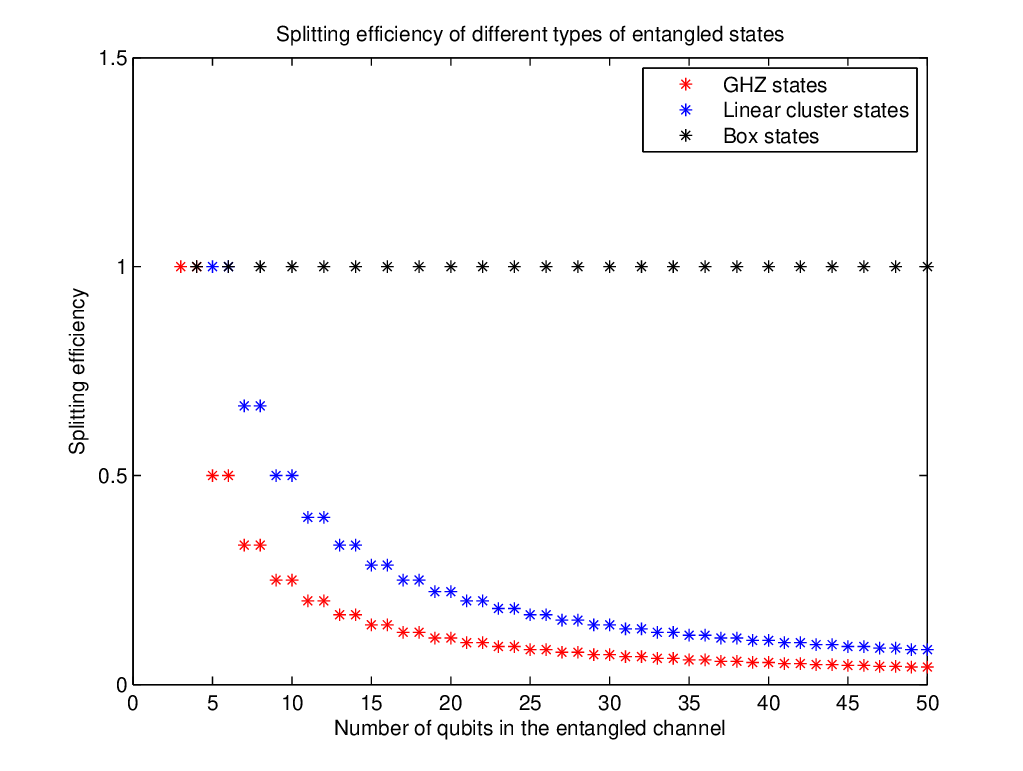} 
\includegraphics[height=1.5in,width=2in,angle=0]{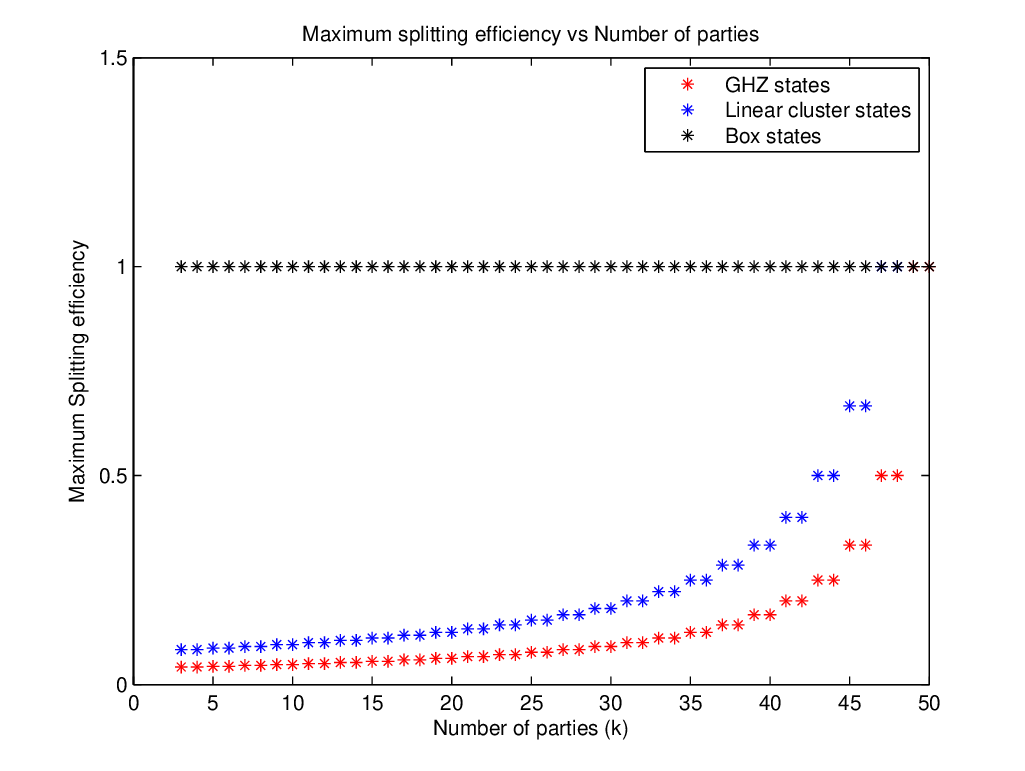} 
\caption{\small \sl The variation of splitting capacity with a) the number of qubits for three parties, b) the number of parties  } 
\end{figure} 
\textit{Conclusions- }We have established a formal understanding of the properties of a physical system which mediates an information splitting protocol
and quantified the number of ways of achieving the same. The study is applicable to a wide range of systems which need not be maximally entangled (as given by Theorem 2)
and is naturally applicable for different kinds of quantum networking protocols that relies on teleportation.  Subsequently, the study 
allows the characterization of multi particle entangled states in terms of its `information splitting efficiency'. 
This result helps us to design an optimal quantum network with minimal resources and 
paves a way towards the  understanding of the usefulness of higher-qubit entangled states for quantum networking protocols. 

\end{document}